\documentclass[twocolumn,showpacs,showkeys,preprintnumbers,
amsmath,amssymb,aps,floatfix,pre]{revtex4}
 
\usepackage{graphicx}
\usepackage{amsmath}
\usepackage{multirow}
\usepackage{times}
\usepackage{amssymb}
\usepackage{dcolumn}%
\usepackage{bm}%
\usepackage{hyperref}
\usepackage{ctable}
\usepackage{tabularx}

\hypersetup
{
	colorlinks=true,
	linkcolor=blue,
	citecolor=blue
}

\begin{document}

\title[]{{Thermodynamic scaling of  vibrational dynamics and relaxation}}
\author{F.\@ Puosi  \footnote{Present address: Sciences et Ing\'enierie des Mat\'eriaux et Proc\'ed\'es (SIMaP), UMR CNRS 5266, Grenoble INP, UGA 1130 Rue de la Piscine, BP 75, 38402 Saint-Martin d'H\`eres Cedex, France}}
\affiliation{Dipartimento di Fisica ``Enrico Fermi'', 
Universit\`a di Pisa, Largo B.\@Pontecorvo 3, I-56127 Pisa, Italy}
 \author{O.\@ Chulkin  \footnote{Present address: Odessa National Polytechnic University, 65044, Odessa, Ukraine.}}
\affiliation{Dipartimento di Fisica ``Enrico Fermi'', 
Universit\`a di Pisa, Largo B.\@Pontecorvo 3, I-56127 Pisa, Italy}
 \author{S.\@ Bernini  \footnote{Present 
 address: Jawaharlal Nehru Center for Advanced Scientific Research, Theoretical Sciences Unit, Jakkur Campus, Bengaluru 560064, India.}}
\affiliation{Dipartimento di Fisica ``Enrico Fermi'', 
Universit\`a di Pisa, Largo B.\@Pontecorvo 3, I-56127 Pisa, Italy}
\author{S.\@ Capaccioli}
\affiliation{Dipartimento di Fisica ``Enrico Fermi'', 
Universit\`a di Pisa, Largo B.\@Pontecorvo 3, I-56127 Pisa, Italy}
\affiliation{IPCF-CNR, UOS Pisa, Italy}
\author{D.\@ Leporini}
\email{dino.leporini@unipi.it}
\affiliation{Dipartimento di Fisica ``Enrico Fermi'', 
Universit\`a di Pisa, Largo B.\@Pontecorvo 3, I-56127 Pisa, Italy}
\affiliation{IPCF-CNR, UOS Pisa, Italy}

\date{\today}

\begin{abstract}
\noindent 
{ We investigate by thorough Molecular Dynamics simulations the thermodynamic scaling (TS)  of a polymer melt. Two distinct models, with strong and weak virial-energy correlations, are considered. Both evidence
the joint TS with the same characteristic exponent $\gamma_{ts}$ of the fast mobility - the mean square amplitude of the picosecond rattling motion inside the cage -, and the much slower structural relaxation and chain reorientation. 
If the cage effect is appreciable, the TS master curves of the fast mobility are nearly linear, grouping in a bundle of approximately concurrent lines for different fragilities.
An expression  of the TS master curve of the structural relaxation with one adjustable parameter less than the available three-parameters alternatives is derived.
The novel expression fits well with the experimental TS master curves of thirty-four glassformers and, in particular, their slope at the glass transition, i.e. the isochoric fragility. For the glassformer OTP  the isochoric fragility allows to satisfactorily predict  the TS master curve of the fast mobility with no adjustments.
}  

\end{abstract}

\maketitle
\section{Introduction}
\label{intro}

Understanding the structural arrest of a supercooled liquid leading to the glass formation is a major scientific challenge in condensed matter physics \cite{BerthierBiroliRMP11,EdigerHarrowell12}.  A remarkable development in understanding the relaxation and the transport of liquids and polymer melts was the discovery that the temperature ($T$) and the density ($\rho$) dependence of, e.g., the structural relaxation time $\tau_\alpha$ and the viscosity $\eta$, can be scaled to a material-dependent master curve \cite{Hollander2001,casaliniPRE04,dreyfusEPJB04,simionescoEPL04,casaliniRepProg05,
RolandReviewMM07,CasaliniRolandPRL14}:
\begin{equation}\label{eqn:tvgammdef}
\log \tau_\alpha, \log \eta = \mathcal{F}_{TS}(T \rho^{-\gamma_{ts}})
\end{equation}
In Eq.\ref{eqn:tvgammdef} both the form of the master curve $\mathcal{F}$ and the exponent $\gamma_{ts}$ are system-specific. The above scaling is usually referred to as "temperature-density scaling" or "thermodynamic scaling" (TS). TS applies to van der Waals liquids, polymers, ionic liquids \cite{casaliniRepProg05,RolandReviewMM07,CasaliniRolandPRL14,rolandJCP06,lopezJCP11,paluchJPCL10}, liquid crystals \cite{RolandLiqCryst11}  and plastic crystals \cite{CapacciPlasticCry16} 
 but not to all of the hydrogen-bonded liquids since the equilibrium structure of the liquid  and its degree of hydrogen bonding are expected to change when temperature and pressure are changed \cite{rolandPRBhb08}.
Regarding network-bonded inorganic glass formers such as silica glasses, from the experimental and numerical studies it seems that the relation of Eq.\ref{eqn:tvgammdef} keeps only locally, i.e. over limited T-P ranges, and the exponent describing the density scaling varies with temperature and volume in a non monotonic way, due to changes in the local environment of the bonded atoms \cite{ShellDebe02,Dreyfus07,McMillan09}.  In general, the scaling exponent $\gamma_{ts}$, which is a measure of the contribution of density relative to that of temperature, varies in the range from $0.13$  to $8.5$ \cite{casaliniRepProg05}. 

TS is attractive for encompassing the changes of both temperature {\it and} density  so that it represents a severe test of theory and models of the structural arrest occurring at the glass transition (GT).
Among the possible justifications of TS, one hypothesis is that the scaling exponent $\gamma_{ts}$ is strictly related to the intermolecular potential. Indeed, for a liquid having a pairwise additive intermolecular potential described by an inverse power law (IPL) $v(r)\propto r^{-n}$, all the reduced thermodynamic and dynamic properties can be expressed in terms of the variable $\rho^{n/3}/T$ \cite{HooverCP71}. The conformance of real materials to TS may result from their intermolecular potential being approximated
 by an IPL, at least in some definite range of intermolecular distance, and consideration of certain dynamic properties \cite{coslovichJPCBlett08}. 
On a more general ground, Dyre and coworkers proved that liquids with strong correlation of the fluctuations of the virial pressure (W) and the potential energy (U), the so called strongly correlating liquids, exhibit TS { and $3 \gamma_{ts}$ is interpreted as the $n$ exponent of an effective IPL potential} \cite{DyreTvGammaJCP09,schroderPRE09,DyreReviewHiddenScaleJPCB14}. Even if sufficient, strong virial-energy correlations are not necessary for TS. Indeed, TS is
observed in experiments concerning a few hydrogen-bonded liquids (e.g. glycerol and sorbitol) \cite{casaliniRepProg05}  { and molecular-dynamics (MD) simulations of supercooled metallic liquids \cite{Hu_2016}. All these systems are not strongly correlating liquids since glassformers with {\it competing} interactions have poor virial-energy correlations \cite{Isom1,CoslovichJNCS11}.}
Competing interactions are also present in {\it molecular} systems where distinct bonding and non-bonding interactions are present. In particular, nearly missing virial-energy correlations have been reported in polymer models made of linear chains with stiff, but not rigid, bonds \cite{TesiFrancesco,DyreVirialEnergyHarmonicPolymJCP15}.

The TS master curve  is not the same for different dynamic properties. As to the structural relaxation, an interpretation of the scaling is to consider the $\tau_\alpha(T,\rho)$ dependence as thermally activated with a $V$ dependent activation energy $\tau_\alpha(T,\rho) \sim \exp(E_A(\rho)/T)$ \cite{tarjusJCP04}. Imposing $E_A(\rho)\propto \rho^{\gamma_{ts}}$,  density scaling is recovered, though such a picture is in contrast with the fact $\tau_\alpha$ is not an exponential function of $T\rho^{-\gamma_{ts}}$  \cite{casaliniRepProg05}.  Entropy has been considered to better understand TS. Casalini \textit{et al.}  used the  entropy model of Avramov \cite{avramovJNCS05} to derive an expression of the relaxation time in terms of the pressure and the temperature which accurately fits the experimental data of several glass-forming liquids and polymers with two adjustable parameters, having taken $\gamma_{ts}$ from the experiment and using $\tau_\alpha(T_g) = 10^2$ s \cite{CasaliniAvra06,CasaliniJNCS07}. Another expression, with the same number of adjustable parameters, based on an entropic model recently formulated by Mauro et al. \cite{MauroPNAS09} has been investigated \cite{PaluchSokolovTvgammaMauroJPCLett12}.

TS has been mostly investigated on the time scale of the structural relaxation or viscous flow. Is it also observed on shorter timescales ? Here we address the picosecond rattling motion, with mean square amplitude $\langle u^2 \rangle$, of the particles trapped by the cage of their neighbours. Henceforth $\langle u^2 \rangle$, which is strictly related to the familiar Debye-Waller factor, will be referred to as "fast mobility". { It is worth noting that, since in viscous liquids the relaxation times $\tau_\alpha$ fairly exceeds the picosecond timescale,  at any given moment of time the fraction of particles undergoing vibrational motion $\phi_{vib}$ is large, $\phi_{vib} \sim 1 -  (\omega_D \tau_\alpha)^{-1} \sim 1$, where $\omega_D \sim 10^{13}$ rad/s is the Debye frequency \cite{TrachenkoRepProgrPhys16}.}
Rattling, a manifestation of the vibrational dynamics, occurs in a soft cage so that the fast mobility is in principle affected by both local aspects, like cage geometry or local rearrangements, as well as extended collective properties like elasticity 
\cite{TrachenkoRepProgrPhys16,PuosiLepoJCPCor12,PuosiLepoJCPCor12_Erratum,BerniniAnisotropSolidLikeJCP16,VoroBinarieJCP15,VoronoiBarcellonaJNCS14,TakeshiDisplCorrFuncPRE16}.
The temporary trapping and subsequent escape mechanisms lead to large fluctuations around the averaged dynamical behavior with strong heterogeneous dynamics \cite{BerthierBiroliRMP11,EdigerHarrowell12} and 
{non-exponential relaxation \cite{MonthBouch96}}.
{The presence of rattling and escape processes in liquids and relationships between them were
first proposed by Maxwell \cite{MaxwellViscoEla1867} and Frenkel  \cite{Frenkel26,FrenkelNature35,Frenkel}, see a recent review \cite{TrachenkoRepProgrPhys16}. Other early} investigations \cite{TobolskyEtAl43,RahmanCageCentroidJCP66}  and recent theoretical \cite{HallWoly87,Angell95,MarAngell01,Ngai00,NgaiAnharmJPCM00,Ngai04,Dyre06,DouglasCiceroneSoftMatter12,WyartPNAS2013,SchweizerElastic1JCP14,SchweizerElastic2JCP14,ElasticoEPJE15} 
 studies addressed the  rattling process in the cage to understand the structural relaxation - the escape process -  gaining support from numerical \cite{Angell68,Nemilov68,Angell95B,StarrEtAl02,BordatNgai04,Harrowell06,Harrowell_NP08,HarrowellJCP09,douglasPNAS09,XiaWolynes00,DudowiczEtAl08,DouglasCiceroneSoftMatter12,OurNatPhys,lepoJCP09,Puosi11,SpecialIssueJCP13,UnivSoftMatter11,CommentSoftMatter13,DouglasStarrPNAS2015,DouglasLocalMod16,OttochianLepoJNCS11,UnivPhilMag11,Puosi12SE,Puosi12,PuosiLepoJCPCor12,PuosiLepoJCPCor12_Erratum,ElasticoEPJE15}  and experimental 
 \cite{BuchZorn92,AndreozziEtAl98,Cicerone11,SokolovNovikovPRL13} works on glassforming liquids. In particular, the role of vibrational anharmonicity as key ingredient of the relaxation has been noted \cite{BordatNgai04,Ngai00,NgaiAnharmJPCM00,NgaiBook}.

Renewed interest about the fast mobility was raised by extensive MD simulations evidencing the universal correlation between the structural relaxation time $\tau_\alpha$ and $\langle u^2\rangle$. 
{ Insight into the correlation is offered by the remark that the height of the barrier to be surmounted for structure rearrangement  increases with the curvature near the minimum of the potential well temporarily trapping the particles, as first noted by Tobolsky et al \cite{TobolskyEtAl43} via a simple viscoelastic model and put on a firmer ground by Hall and Wolynes who related the barrier height to $1/ \langle u^2\rangle$ \cite{HallWoly87}.}
The correlation was reported in polymeric systems \cite{OurNatPhys,lepoJCP09,Puosi11}, binary atomic mixtures \cite{lepoJCP09,SpecialIssueJCP13,VoroBinarieJCP15,DouglasLocalMod16}, colloidal gels \cite{UnivSoftMatter11} and antiplasticized polymers \cite{DouglasCiceroneSoftMatter12,DouglasStarrPNAS2015}
and compared with the experimental data concerning several glassformers in  a wide range of fragility - the steepness $m$ of the temperature-dependence of the logarithm of the structural relaxation time at GT defined by Angell \cite{Angell91} - ($20 \le m \le 191$), including polymers, van der Waals and hydrogen-bonded liquids, metallic glasses, molten salts and the strongest inorganic glassformers \cite{OurNatPhys,UnivPhilMag11,OttochianLepoJNCS11,SpecialIssueJCP13,CommentSoftMatter13,SokolovNovikovPRL13}. The correlation between structural relaxation and fast mobility is summarized by  the universal master curve \cite{OurNatPhys}:
\begin{eqnarray}
\label{eqn:u2tauExp0}
\log \tau_\alpha &=& \mathcal{F}_{FM}(\langle u^2\rangle) \\ 
\label{eqn:u2tauExp}
&=& \alpha+\tilde{\beta}\left (\frac{\langle u_g^2\rangle}{\langle u^2\rangle} \right)+\tilde{\gamma}\left (\frac{\langle u_g^2\rangle}{\langle u^2\rangle} \right)^2
\end{eqnarray}
$ \langle u^2_g\rangle$ is the fast mobility at GT,
$\tilde{\beta}$ and $\tilde{\gamma}$ are suitable universal constants independent of the kinetic fragility \cite{OurNatPhys,SpecialIssueJCP13}, and $\alpha =  2 - \tilde{\beta} - \tilde{\gamma}$ to comply with the usual definition $\tau_\alpha = 100$ s at the glass transition. 
Therefore, it is noteworthy that a different definition for the timescale related to the GT modifies the expression of Eq.\ref{eqn:u2tauExp} only by shifting for a constant value.
Eq.\ref{eqn:u2tauExp} has been tested on experimental data \cite{OurNatPhys,OttochianLepoJNCS11,UnivPhilMag11,SpecialIssueJCP13,CommentSoftMatter13,SokolovNovikovPRL13} as well as numerical models of polymers \cite{OurNatPhys,lepoJCP09,Puosi11,Puosi12SE,VoroBinarieJCP15,BerniniAnisotropSolidLikeJCP16}, colloids \cite{UnivSoftMatter11} and atomic liquids \cite{lepoJCP09,SpecialIssueJCP13,VoroBinarieJCP15}. Douglas and coworkers  developed a localization model predicting the alternative master curve $\mathcal{F}_{FM}(\langle u^2\rangle) \propto \langle u^2\rangle^{-3/2}$ relating the structural relaxation time and the fast mobility  \cite{DouglasCiceroneSoftMatter12,DouglasStarrPNAS2015,DouglasLocalMod16}. Both the latter form and Eq.\ref{eqn:u2tauExp} account for the convexity of the master curve, evidenced by the experiments and simulations, and improve the original linear relation proposed by Hall and Wolynes in their pioneering work \cite{HallWoly87}.

We carry out a detailed study of TS of, jointly, the fast dynamics - as sensed by  the fast mobility - and the much slower structural relaxation and chain reorientation. The matter is investigated by MD simulations of a coarse-grained polymer model and comparison with the available experimental data. { In the MD study the polymer chain is modelled  with either rigid or semi-rigid bonds. The variant of the polymer model with semi-rigid bonds, differently from the one with rigid bonds \cite{DyreVirialEnergyRigidPolymJCP14}, is not a strongly-correlating liquid, as previously noted \cite{TesiFrancesco} and recently reported \cite{DyreVirialEnergyHarmonicPolymJCP15} for a closely related model, owing to the competition between the bonding and the non-bonding interactions \cite{Isom1,CoslovichJNCS11}. This means that there is no effective inverse-power law potential replacing the actual particle-particle interaction potential, thus precluding the usual TS interpretation. }

TS of the fast mobility of the molten salt CKN \cite{ruoccoJCP11u2}, polymers \cite{TesiChulkin} and binary atomic mixtures \cite{SpecialIssueJCP13} has been reported by previous MD studies. 

The paper is organized as follows: Sec.\ref{numerical} gives details about the MD simulations, Sec.\ref{ResDisc} presents the results of the MD simulations and the comparison with the experimental data. Finally, Sec.\ref{conclu} summarizes the conclusions.

\begin{figure}[t]
  \begin{center}
    \includegraphics[width=0.9\linewidth]{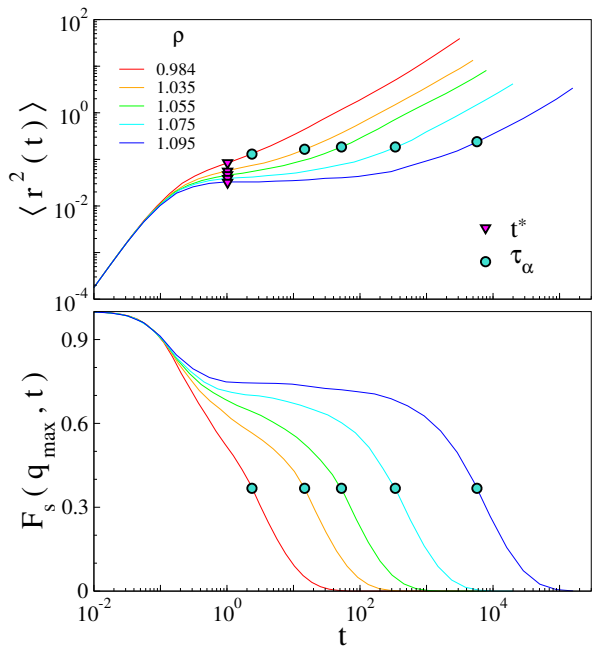}
    \caption{\label{FigIntro} Monomer MSD, Eq.\ref{Eq:MSD} (top) and corresponding ISF, Eq.\ref{Eq:Fself} (bottom) { from MD simulations} of the SB trimers with LJ non-bonding potential at $T=0.6$ and different densities. The triangles and dots mark the positions of the inflection point in the MSD ( $t^*$ ) and the relaxation time ( $\tau_\alpha$ ), respectively. { All the quantities are in reduced MD units.}}
  \end{center}
\end{figure}

\section{Methods }
\label{numerical}

A coarse-grained model of a melt of linear fully-flexible unentangled polymer chains with $M$ monomers each is used. The chains are fully-flexible, i.e. bond-bending and bond-torsions potentials are not present. The system has $N=2000$ monomers in all cases but $M=3$, where $N=2001$. Nonbonded monomers at a distance $r$  interact via the truncated Mie potential \cite{VoroBinarieJCP15}:
\begin{equation}
U_{q,p}(r)=\frac{\epsilon}{q-p}\left [ p\left (\frac{\sigma^*}{r}\right)^q - q\left (\frac{\sigma^*}{r}\right)^p \right]+U_{cut}
\label{mie}
\end{equation}
for $r<r_c=2.5\sigma$ and zero otherwise, where $U_{q,p}(r)=U_{p,q}(r)$ and   
$\sigma^*=2^{1/6}\sigma$ is the position of the potential minimum with depth $\epsilon$. The value of the constant $U_{cut}$ is chosen to 
ensure that $U_{q,p}(r)$ is continuous at $r=r_c$. $U_{6,12}$ is the usual 
Lennard-Jones (LJ) potential. Changing the $p$ and $q$ 
parameters does not affect the position  $r=\sigma^*$ or the depth $\epsilon$ 
of the potential minimum but only the steepness of the repulsive and the 
attractive wings, see Fig.\ref{fragility}. We varied the number density $\rho$, 
the temperature $T$, the chain length and the $(p,q)$ parameters of the 
non-bonding potential $U_{q,p}(r)$, Eq.\ref{mie}.   In particular, we changed the $q$ parameter with the prescription $q > p = 6$, i.e. we modelled the attractive tail by the 
 London dispersion interaction and varied the steepness of the repulsive part \cite{JacksonMiePolymerJCP13}.
Two different kinds of bonding are also considered. In the case of semi-rigid bonds (SB) bonded monomers interact with a potential which is the sum of the finitely extendible nonlinear elastic (FENE) potential and the LJ potential \cite{sim}. The resulting bond lenght is $b=0.97\,\sigma$ within few percent. 
Alternatively, in the case of rigid bonds (RB) bonded monomers are constrained 
to a distance $b=0.97\,\sigma$ by using the RATTLE algorithm 
\cite{allentildesley}. All the $\sim 230$ states are simulated and listed in Appendix A2 of 
ref. \cite{TesiFrancesco}. With the purpose of plotting 
Fig.\ref{fig:u2Tvglong}, the melt of trimers with non-bonding LJ potential, 
$(p,q)=(6,12)$ is also studied at the following densities and temperatures 
[$\rho$;$T_1$,$T_2$, ...]: [1;1,1.2,1.4,1.6, 1.8, 2, 
2.2, 2.4, 2.6, 2.8, 3], [0.95; 3], [0.9; 3], [0.85; 2, 3], [1.05; 1.3, 2, 2.65, 
3.3, 3.98, 4.65, 5.3], [1.025; 1.44, 2.02, 2.6, 3.17, 3.75, 4.33, 4.9].

All quantities are in reduced units: length in units of $\sigma$, temperature 
in units of $\epsilon/k_B$ and time in units of $\sigma\sqrt{\mu/\epsilon}$ 
where $\mu$ is the monomer mass \cite{MauriLepEPL06}. We set $\mu=k_B=1$. { It is interesting to map the reduced MD units to real physical units.  As an example for polyethylene and polystyrene it was found
$\sigma=5.3$ \AA, $\varepsilon/k_B = 443$ K ,$\tau_{MD} = 1.8$͒ ps and $\sigma=9.7$ \AA, $\varepsilon/k_B = 490$ K ,$\tau_{MD} = 9$ ps, respectively \cite{Kroger04}. }

$NPT$ and $NTV$ ensembles have been used for equilibration runs, while $NVE$ 
ensemble has been used for production runs for a given state point. $NPT$ and 
$NTV$ ensembles are studied by the extended system method introduced by 
Andersen \cite{Andersen80} and Nos\'e \cite{NTVnose}. The numerical integration 
of the augmented Hamiltonian is performed through the multiple time steps 
algorithm, reversible Reference System Propagator Algotithm (r-RESPA), 
developed by Tuckerman \textit{et al} \cite{respa}. In particular, the $NPT$ 
and $NTV$  operators is factorized using the Trotter theorem \cite{trotter} 
separating the short range and long range contributions of the potential, 
according to the Weeks-Chandler-Andersen (WCA) decomposition \cite{tuckWCA}. 
Other details are given elsewhere \cite{OurNatPhys,Puosi11,lepoJCP09}.

\section{Results and discussion}
\label{ResDisc}

\subsection{General aspects}
\label{general}

\subsubsection{Mobility and relaxation}
\label{MobilRelaxElastic}

We define the monomer displacement in a time $t$ as:
\begin{equation}
\Delta \mathbf{r}_i (t) = \mathbf{r}_i(t)-\mathbf{r}_i(0)
\label{mondispl} 
\end{equation}
where $\mathbf{r}_i(t)$ is the vector position of the $i$-th monomer at time $t$.
The mean square displacement (MSD) $\langle r^2(t)\rangle$ is expressed as:
\begin{equation}
\langle r^2(t)\rangle = \left\langle \frac{1}{N} \sum_{i=1}^N  \|\Delta 
\mathbf{r}_i (t)\|^2 \right \rangle
\label{Eq:MSD}  
\end{equation}
where brackets denote the ensemble average. In addition to MSD the incoherent, self part of the intermediate scattering function (ISF) is also considered:
\begin{equation}
F_{s}(q,t) = \left\langle \frac{1}{N} \sum_{j=1}^N e^{i{\bf q}\cdot \Delta 
\mathbf{r}_j (t)} \right\rangle
\label{Eq:Fself} 
\end{equation}
ISF was evaluated at $q= q_{max} $, the maximum of the static structure factor ( $7.06 \le q_{max}\le 7.35$  ).

\begin{figure}[t]
  \begin{center}
    \includegraphics[width=0.99\linewidth]{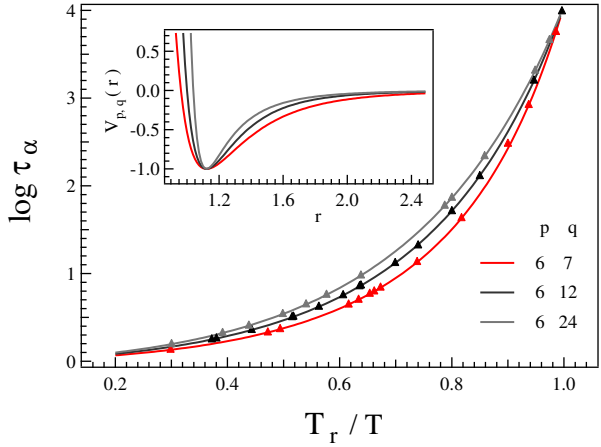}
    \caption{\label{fragility} Reduced temperature dependence of the relaxation time { from MD simulations} of SB trimers ($M=3$) with different forms of the interaction potential between non-bonded monomers, Eq.\ref{mie} (insert). $T_r$ is the temperature where $\tau_\alpha = 10^4$. The number density is $\rho = 1.033$. The fragility increases by decreasing the steepness of the potential around the minimum. The lines are Eq.\ref{eqn:tauTvgMD} with parameters as given by Table \ref{tab:u2tvg} and $\alpha =-0.424(1), \beta = 2.7(1) \cdot 10^{-2}, \gamma =  3.41(3) \cdot 10^{-3}$  \cite{OurNatPhys}. { All the quantities are in reduced MD units.}}
  \end{center}
\end{figure}

Fig.\ref{FigIntro} shows illustrative examples of the monomer MSD (top) and ISF (bottom) curves for states at temperature $T=0.6$ and different densities. At very short times (ballistic regime) MSD increases according to $\langle r^2(t)\rangle \cong (3 k_B T/m) t^2$ and ISF starts to decay. 
The repeated collisions slow the displacement of the tagged monomer, as evinced by the knee of MSD at $ t \sim t_m = 0.175$ which is very close to the minimum of the velocity correlation function  \cite{BerniniAnisotropSolidLikeJCP16}. 
At later times, when the temperature is lowered and/or the density is increased, a quasi-plateau region occurs in both MSD and ISF, and an inflection point is seen at $t^*\simeq 1.023$ in the log-log MSD plot, see Fig.\ref{FigIntro} (top) and, for more details, Ref. \cite{OurNatPhys}. 
The time $t^*$ has been interpreted as the fast $\beta$-relaxation time scale, as described by Mode Coupling Theory
 \cite{SastryPRL16}. 
$t^*$ is state-independent in the present model \cite{OurNatPhys}. The inflection point signals the end of the exploration of the cage by the trapped particle and the subsequent early escapes. 
We define the fast mobility of the monomers of the linear chains  as the MSD at $t^*$  \cite{OurNatPhys}:
\begin{equation}
 \langle u^2 \rangle = \langle r^2(t=t^*)\rangle  \label{ST-MSD}
\end{equation}
The fast mobility is the mean square amplitude of the position fluctuations of the tagged particle in the cage of the neighbours.   The inflection point in the log-log MSD plot disappears if $\langle u^2 \rangle > \langle u^2_m \rangle  = 0.125$ signalling the absence of significant cage effect by the neighbours of the tagged particle \cite{OurNatPhys}. The structural relaxation time $\tau_{\alpha}$, the average escape time from the cage, is defined by the relation $F_s(q_{max}, \tau_{\alpha}) = e^{-1}$. For $t  > \tau_{\alpha}$ MSD increases more steeply and finally ends up in the diffusive regime, whereas ISF decays to zero as a stretched exponential with stretching parameter $\beta \sim 0.6 $ \cite{BerniniCosTetaJPCM2016}.

Fig.\ref{fragility} { presents the temperature dependence of the structural relaxation of the SB trimers for a given density. Data are presented as an Angell plot \cite{Angell91} in terms of the reference temperature $T_r$ where $\tau_\alpha = 10^4$ in MD units, corresponding to about $10-100$ ns.  The plot} illustrates  the changes of the fragility resulting from the different choices of the nonbonding potential. 
We remind that fragility is a measure of the steepness of the temperature-dependence of the logarithm of the structural relaxation time on approaching GT \cite{Angell91}.
It is seen that { more gradual} potentials, giving origin to broader energy minima, associate to higher fragility, as already noted \cite{BordatNgai04}. 

\begin{figure}[t]
  \begin{center}
    \includegraphics[width=0.99\linewidth]{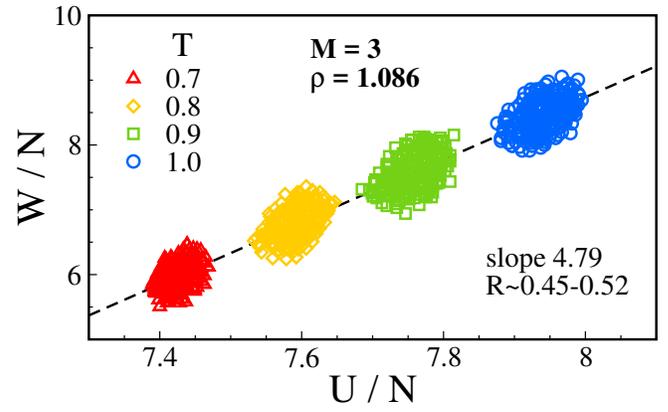}
  \caption[uno]{\label{fig:expIPL}  { Correlation plot of the virial and the configurational energy per particle
   from MD simulations of SB trimers with LJ non-bonding potential
   for states with same density and different temperatures. All the quantities are in reduced MD units.}}
  \end{center}
\end{figure}

{
\subsubsection{Virial-energy correlations}
\label{ThermoVsVirial}

In the case of pair potentials, the virial $W$, i.e., the configurational contribution to pressure, is given by \cite{allentildesley}:
\begin{equation}\label{eqn:virial}
W=-\frac{1}{3}\sum_{i>j}w(|{\bf r}_i-{\bf r}_j |) 
\end{equation}
where  $w(r)=rv^{\prime}(r)$, $v^{\prime}$ being the derivative of the pair potential $v(r)$. For an IPL potential, $v(r)\propto r^{-n}$, one has $w(r)=-n \, v(r)$ and the virial is proportional to the potential energy $U=\sum_{i>j}v(|{\bf r}_i-{\bf r}_j |)$:
\begin{equation}\label{eqn:virpotipl}
W=\frac{n}{3}U
\end{equation}
Eq.\ref{eqn:virpotipl} states that in IPL systems, irrespective of the physical state, the scatter plot of the instantaneous potential energy and virial shows perfect correlation with slope $n/3$. Liquids with strong virial-energy correlations exhibits TS with $\gamma_{ts} = n/3$ \cite{DyreTvGammaJCP09,schroderPRE09,DyreReviewHiddenScaleJPCB14}.
Figure \ref{fig:expIPL} plots the instantaneous virial and potential energy fluctuations of  SB trimers with non-bonding LJ potential. The degree of correlation is quantified by the correlation coefficient $R$:
\begin{equation}\label{eqn:corcoef}
R=\frac{\langle \Delta W  \Delta U \rangle}{\sqrt{\langle (\Delta W)^2 \rangle}\sqrt{\langle (\Delta U)^2 \rangle}}
\end{equation}
where $\Delta$ denotes the deviation from the average value of the given quantity and $\langle ... \rangle$ denotes the thermal averages. We find low correlation, $R\sim0.45-0.52$, depending on the state. Differently, in the case of RB chains the correlation is high, $R > 0.8$ (not shown), as in previous studies on linear chains with rigid bonds \cite{TesiFrancesco,DyreVirialEnergyRigidPolymJCP14}. The drop of the virial-energy correlations by replacing rigid bonds with semirigid ones in linear chains has been noted \cite{TesiFrancesco} and recently reported \cite{DyreVirialEnergyHarmonicPolymJCP15} and is ascribed to the competition between the bonding and the non-bonding interactions \cite{Isom1,CoslovichJNCS11}. 
}

\begin{figure}[t]
  \begin{center}
    \includegraphics[width=0.75\linewidth]{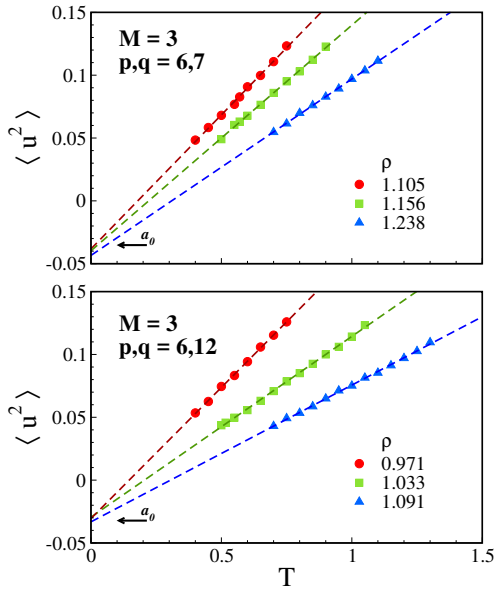}
    \caption[uno]{\label{fig:u2T} \small{Temperature dependence of the fast mobility $\langle u^2 \rangle$ along different isochores, from MD simulations of SB trimers and non-bonding potential, Eq.\ref{mie},  with $p,q=6,7$ (top) and $p,q=6,12$ (bottom). The dashed lines are the best-fit curves according to Eq.\ref{eqn:u2Tisoc}. Their extrapolation to $T\rightarrow 0 $ gives the parameter $a_0$. { All the quantities are in reduced MD units.} }}
  \end{center}
\end{figure}

\begin{figure}[t]\label{fig:linu2isot}
  \begin{center}
    \includegraphics[width=0.75\linewidth]{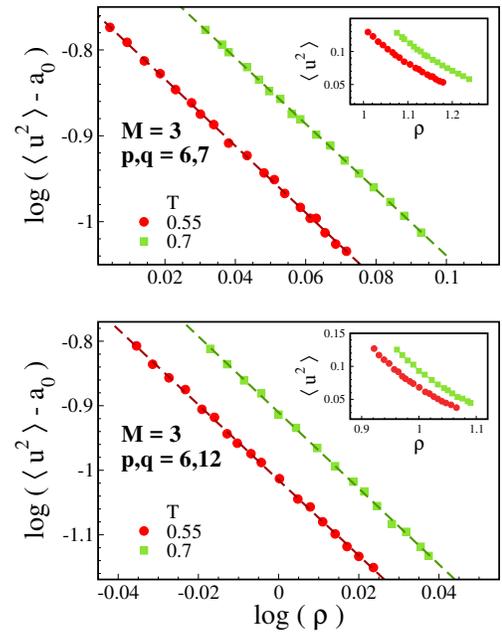}
    \caption[uno]{\label{fig:u2rho} \small{Density-dependence of the quantity $\langle u^2\rangle-a_0$, see Eq.\ref{eqn:u2Tisoc}, along different isotherms for the same systems of Fig. \ref{fig:u2T}. The slope of the best-fit lines (dashed line) gives the exponent of the power-law dependence on density. Insets:
    the fast mobility $\langle u^2\rangle$ versus the density for the same state points of the main panels.  { All the quantities are in reduced MD units.}}}
  \end{center}
\end{figure}

\begin{table}[b]
\begin{center}
 \begin{tabular}{| l | c | c | c | c | c | c | }
\hline	
  {} & {Bonds} 	&  $M$ 	& $(p,q) $ 	& $\gamma_{ts}$ 	& $a_0$  	& $a_1$  \\
\hline
  A & SB & $ 3$ & $ (6,7)$ & $ 3.9(1)$ & $ -0.039(2)$ & $ 0.317(5)$ \\
  B & SB & $ 10$ & $ (6,8)$  & $ 4.7(2)$ & $ -0.036(2)$ & $ 0.283(6)$ \\
  C & SB & $ 3$ & $ (6,8)$  &$ 4.3(1)$ & $ -0.037(1)$ & $ 0.279(4)$ \\
  D & SB & $ 10$ & $ (6,10)$  & $ 5.9(2)$ & $ -0.032(1)$ & $ 0.229(5)$ \\
  E & SB & $ 3$ & $ (6,10)$  &$ 5.2.(1)$ & $ -0.040(1)$ & $ 0.244(6)$ \\
  F & SB & $ 10$ & $ (6,12)$  & $ 6.7(1)$ & $ -0.022(1)$ & $ 0.170(4)$ \\
  G & RB & $ 10$ & $ (6,12)$  & $ 6.65(5)$ & $-0.029(1)$ & $ 0.162(5)$ \\
  H & SB & $ 3$ & $ (6,12)$   & $ 5.80(1)$ & $ -0.029(1)$ & $ 0.172(4)$ \\
  I & RB & $ 3$ & $ (6,12)$  & $ 5.85(5)$ & $ -0.033(2)$ & $ 0.169(5)$ \\
  L &SB & $ 3$ & $ (6,18)$  & $ 7.6(2)$ & $-0.029(2)$ & $ 0.110(5)$ \\
  M & SB & $ 3$ & $ (6,24)$  &  $ 8.4(2)$ & $ -0.023(1)$ & $ 0.074(5)$ \\
\hline 

 \end{tabular}
\caption{ \label{tab:u2tvg} {The density scaling exponent $\gamma_{ts}$  and the parameters $a_0$ and $a_1$ of eq \ref{eqn:u2tvg} for the systems of  Figures \ref{fig:u2TvGammaP6} and \ref{fig:u2TvGamma}. }
}
\end{center}
\end{table}

\subsection{Thermodynamic scaling of the fast mobility}
\label{ThermoFastMobil}

\subsubsection{Master curve in the cage regime}
\label{mastercurve}

We now derive the expression of the TS master curve of the fast mobility. 
We start by investigating  the temperature dependence of  $\langle u^2\rangle$. In Fig. \ref{fig:u2T} the temperature behavior of the fast mobility  along different isochores is shown for SB trimers with different non-bonding potential, leading to {\it different fragility}, see Fig. \ref{fragility}. We observe that in the considered temperature range $\langle u^2 \rangle$ shows a well-defined linear variation, which is well fitted by the equation
\begin{equation}\label{eqn:u2Tisoc}
\langle u^2 (T) \rangle= a_0 +  m\cdot T
\end{equation}
where $a_0$ and $m$ are suitable constants. Fig. \ref{fig:u2T} shows that $a_0$ depends very weakly on the density within the errors. Instead, the slope of $\langle u^2(T)\rangle$ is a decreasing function of the density.  Table \ref{tab:u2tvg} lists the $a_0$ best-fit values of all the systems of interest. For a given system, at least two different isochores are used.

\begin{figure}[t]
  \begin{center}
\includegraphics[width=0.99\linewidth]{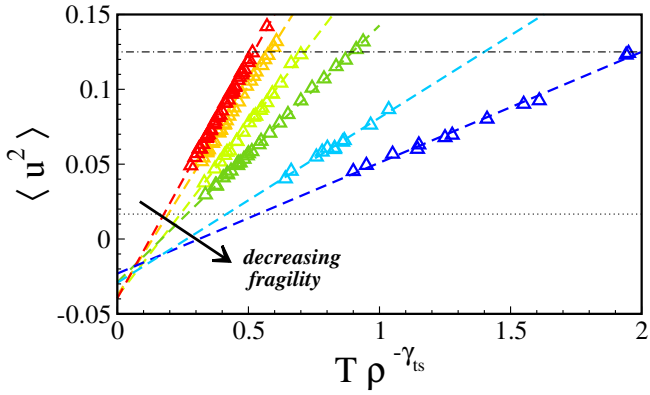}
    \caption[uno]{\label{fig:u2TvGammaP6} \small{TS  of the fast mobility { from MD simulations} of SB trimers with different non-bonding potential, Eq.\ref{mie} ($p=6$, $q=7,8,10,12,18,24$). They exhibit decreasing fragilities with increasing $q$ values, see Fig.\ref{fragility}. 
    The dashed lines are the master curves Eq.\ref{eqn:u2tvg} with parameters listed in Table \ref{tab:u2tvg}. The dotted line marks  $\langle u^2_g \rangle\approx 0.0166$, the fast mobility at the glass transition \cite{OurNatPhys}. The dot-dashed line marks the maximum fast mobility in the presence of caging, $\langle u^2_m \rangle=  0.125$, see Sec.\ref{MobilRelaxElastic}. The extrapolated master curves intersect approximately at $(0.08, -0.02)$. { All the quantities are in reduced MD units.} }}
  \end{center}
\end{figure}

According to Figure \ref{fig:u2T}, the fast mobility $\langle u^2(T)\rangle $ vanishes at the finite temperature $T_{0}^{(FM)}$. Zhang \textit{et al. } showed that $T_{0}^{(FM)}$ coincides, within the uncertainty, with the Vogel-Fulcher-Tammann (VFT) temperature  $T_0$ where the structural relaxation time diverges \cite{douglasPNAS09}. We test this conclusion on the set where the temperature dependence was studied in greatest detail  ($M=3$, $\rho=1.033$ and $p,q=6,12$, see Fig.\ref{fragility}). We  find that the fast mobility tends to vanish at $T_{0}^{(FM)}=0.20(1)$, which is slightly smaller than the VFT temperature $T_0=0.28(2)$, obtained by the best-fit of the corresponding structural relaxation time $\tau_\alpha$.
 It has to be noted that errors can arise from the determination of the VFT temperature $T_0$ evaluating a non-linear function on a higher temperature range of data. In particular, the validity of VFT function well below GT temperature is still matter of debate \cite{McKenna08,DyreNatPhys08}.

We now discuss the density dependence of the fast mobility. From the analysis of Fig.\ref{fig:u2T}, we know that the $\rho$-dependence of the fast mobility is virtually all incorporated in the slope $m$ in Eq.\ref{eqn:u2Tisoc} . Fig. \ref{fig:u2rho} plots the quantity $m \cdot T$ along two isotherms for the systems of Fig.\ref{fig:u2T}.  It is seen that  the slope $m$ exhibits a power-law dependence on density:
\begin{equation}
m =  a_1\cdot \rho^{-\gamma_{ts}}
\label{slope}
\end{equation}
The above procedure involving isochores and isotherms leads to the TS master curve of  the fast mobility:
\begin{equation}
\label{eqn:u2tvg}
\langle u^2 (T,\rho) \rangle= a_0 +  a_1\cdot T\rho^{-\gamma_{ts}}
\end{equation}
where the parameters $a_0$, $a_1$ and $\gamma_{ts}$ depend in general on the chain length, the monomer-monomer non-bonding potential and the nature of the bonding interaction. Table \ref{tab:u2tvg} lists the best-fit values of $a_1$ and $\gamma_{ts}$ for the systems of interest.
{It will be shown in Sec.\ref{mastercurveRes2} that Eq. \ref{eqn:u2tvg} holds true also in the region where the cage effect disappears for high $T\rho^{-\gamma_{ts}}$ values . On the other hand, Eq. \ref{eqn:u2tvg} breaks down for low $T\rho^{-\gamma_{ts}}$ values where it predicts a non-physical negative fast mobility since $a_0 < 0$, see Table \ref{tab:u2tvg}.}

\begin{figure}[t]
  \begin{center}
    \includegraphics[width=0.9\linewidth]{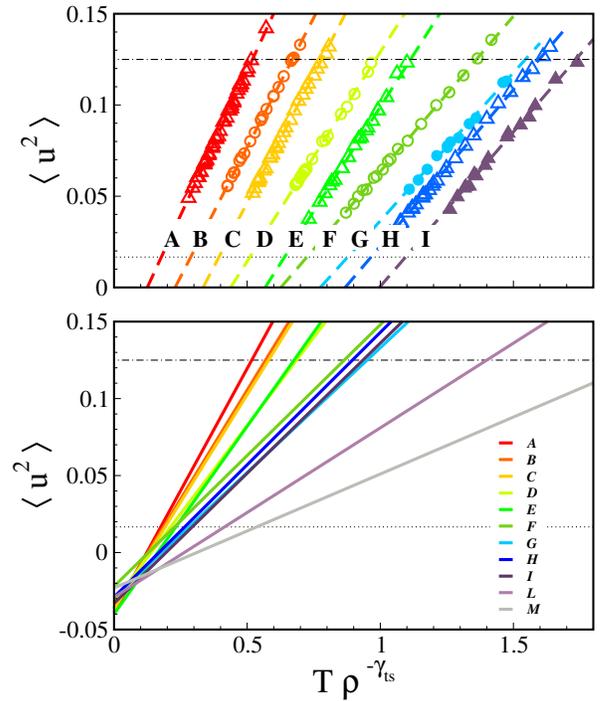}
    \caption[uno]{\label{fig:u2TvGamma} \small{Top: TS of the fast mobility { from MD simulations} of SB and RB polymer chains with different chain length $M$ and  non-bonding potential $U_{q,p}$, Eq.\ref{mie}. For clarity reasons, data are horizontally shifted. A: SB $M=3$, $p,q=6,7$ (shift:$+0.0)$;  B: SB $M=10$, $p,q=6,8$ ($+0.1$); C: SB $M=3$, $p,q=6,8$ ($+0.2$); D: SB $M=10$, $p,q=6,10$ ($+0.3$); E: SB $M=3$, $p,q=6,10$ ($+0.4$); F: SB $M=10$, $p,q=6,12$ ($+0.5$); G: RB $M=10$, $p,q=6,12$ ($+0.6$); H: SB $M=3$, $p,q=6,12$ ($+0.7$); I: RB $M=3$, $p,q=6,12$ ($+0.8$).  The dashed lines are the master curves  Eq.\ref{eqn:u2tvg} with parameters listed in Table \ref{tab:u2tvg}. The dotted line marks  $\langle u^2_g \rangle\approx0.0166$, the fast mobility at the glass transition \cite{OurNatPhys}. The dot-dashed line marks the maximum fast mobility in the presence of caging, $\langle u^2_m \rangle=  0.125$, see Sec.\ref{MobilRelaxElastic}. Bottom: approximated common intersection of all the TS master curves. For clarity reasons MD points are removed. Details about the L and M lines, the two lines with smaller slope in Fig.\ref{fig:u2TvGammaP6}, are given in Table \ref{tab:u2tvg}. { All the quantities are in reduced MD units.}}}
  \end{center}
\end{figure}
\begin{figure}[t]
  \begin{center}
    \includegraphics[width=0.99\linewidth]{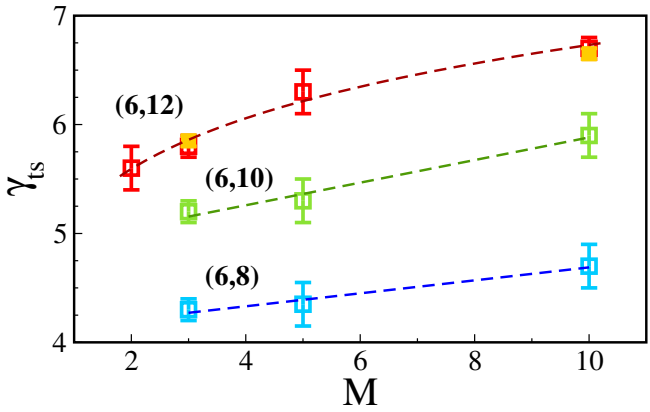}
    \caption[uno]{\label{fig:gammaM} \small{Scaling exponent $\gamma_{ts}$ versus chain length $M$ { from MD simulations of} polymer systems with either SB  (open symbols) or RB  (full symbols) bonds and different non-bonding potential, Eq.\ref{mie}, as indicated by the $(p,q)$ pairs in parenthesis. The dashed lines are guides for the eyes. The exponent $\gamma_{ts}$ increases with: i)  the chain length and, ii) the steepness of the non-bonding potential around the minimum, see Fig.\ref{fragility}.}}
  \end{center}
\end{figure}

\subsubsection{Thermodynamic scaling in the cage regime}
\label{mastercurveRes}

We have extensively investigated TS of the fast mobility in a wide range of different physical states of the systems characterized by the chain length, the bonding and the non-bonding potentials listed in Table \ref{tab:u2tvg}.  The results are presented in Fig. \ref{fig:u2TvGammaP6} and Fig.\ref{fig:u2TvGamma}. We always  find that the procedure outlined in Sec.\ref{mastercurve} leads to a  quite effective TS with master curve nicely fitted by Eq. \ref{eqn:u2tvg}. 

In particular, Fig.\ref{fig:u2TvGammaP6} presents the results concerning SB trimers with non-bonding potential having different steepness and then different fragilities, as suggested by Fig.\ref{fragility}. 
{Fig. \ref{fig:u2TvGammaP6} shows that the linear master curves, Eq. \ref{eqn:u2tvg}, intersect approximately in a single point, thus suggesting that the two parameters $a_0$ and $a_1$ are mutually dependent and, actually, each master curve may be labelled by a {\it single} parameter, e.g. the slope. Given the universal equation Eq.\ref{eqn:u2tauExp}, connecting $\langle u^2 \rangle$ and $\tau_\alpha$,
this implies that TS of the structural relaxation of our melt of trimers is controlled by a {\it single} parameter. By inspecting the results listed in Table \ref{tab:u2tvg}, one finds that the location of the (approximated) intersection depends only mildly on both the chain length and the nature of the bond. { The intersection is rooted in the coupling between the fast dynamics and the anharmonic elasticity. The proof goes fairly beyond the purposes of the present work and will be presented elsewhere \cite{elsewhere}. }
 Fig.\ref{fig:u2TvGamma} (top) plots the TS master curves of a variety of systems with different chain length, non-bonding potentials and bond stiffness, see Table \ref{tab:u2tvg}. It is seen that the linear master curve covers from close to the glass transition up to the boundary of the regime where the cage effect is apparent. {Fig.\ref{fig:u2TvGamma} (bottom) shows that, as already noted in Fig.\ref{fig:u2TvGammaP6}, also in Fig.\ref{fig:u2TvGamma}, the master curves of the fast mobility intersect in a narrow region, so that each master curve may be labelled by a single parameter.}

It seems proper to discuss the main factors affecting the magnitude of the scaling exponent $\gamma_{ts}$. We remind that our linear chains are modelled as {\it fully-flexible}, i.e. bond-bending and bond-torsions potentials are not present. 
Fig.\ref{fig:gammaM} shows that the exponent $\gamma_{ts}$ increases with both the steepness of the non-bonding potential around the minimum, see Fig.\ref{fragility}, - as expected since the approximating IPL potential become stiffer \cite{coslovichJPCBlett08} -, and, mildly, the chain length. 
Since increasing the chain length replaces non-bonding interactions with stiffer bonding ones, we may conclude that in a melt of fully-flexible chains $\gamma_{ts}$ is a measure of the overall stiffness of the system.
The MD simulations of our polymer model yield $\gamma_{ts}$ around $4-7$. 
It is tempting to point out that the polysiloxanes, which, like our model, have a very flexible chain, are characterized by the scaling exponent $\gamma_{ts} \gtrsim 5$, independent of the chain length \cite{casaliniCollPol04}. 
The influence of the chain flexibility on the magnitude of the scaling exponent $\gamma_{ts}$ has distinct features. In stiffer polymers, like polymethylmethacrylate, the exponent decreases abruptly as the length of the chain increases \cite{capacciPMMA}. It must be also pointed out that high molecular weight polymers  are characterized by small values of the exponent, $\gamma_{ts}<3$ \cite{casaliniRepProg05}. These small values are mainly due to the relative stiffness of the chain units, with respect to the mobility corresponding to the  intermolecular degrees of freedom  that are thermally activated: the stiff chain structure hinders rearrangements, resulting in smaller sensitivity to volume effects \cite{capacciPMMA}.
In other terms, adding barriers to intramolecular degrees of freedom of polymers makes the apparent potential softer \cite{rolandJCP06}:  using a proper torsional potential for bonding rotation, for instance an harmonic potential, Tsolou et al. \cite{mavrantzasJCP06}  obtained $\gamma_{ts}$ less than 3 for simulated 1,4-polybutadiene.
On the other hand, our findings are in good agreement with the results obtained on flexible LJ chain fluids by MD simulations and comparison with experiments on some real simple fluids (flexible alkanes), where the scaling exponent was found to vary from 5 to 6.6 on increasing the chain lenghts \cite{Galliero11}.

\begin{figure}[t]
  \begin{center}
    \includegraphics[width=0.8\linewidth]{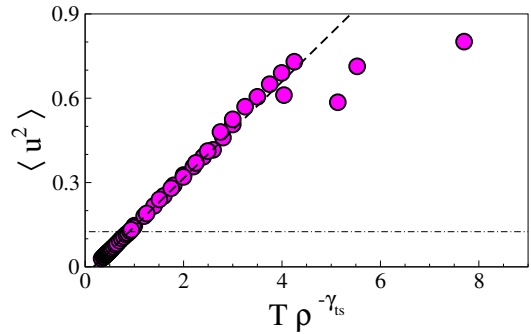}
    \caption[uno]{\label{fig:u2Tvglong} \small{TS of the fast mobilty { from MD simulations} of the trimer with SB bonds and LJ non-bonding potential for high $T\rho^{-\gamma_{ts}}$ values. For $\langle u^2 \rangle > \langle u^2_m \rangle = 0.125$  the cage effect disappears, see Sec.\ref{MobilRelaxElastic}. The linear TS master curve,eq \ref{eqn:u2tvg}, is also drawn (dashed line) with parameters as in Table \ref{tab:u2tvg}. Note that Eq. \ref{eqn:u2tvg} approximates the TS master curve also in part of the region with no cage effect. { All the quantities are in reduced MD units.}}}
  \end{center}
\end{figure}

{ A test of our results on the TS scaling of the fast mobility of polymers is provided by the diffusivity $D$. We know from previous studies on fully-flexible, unentangled polymers \cite{Puosi11,Puosi12SE} and binary atomic mixtures \cite{SpecialIssueJCP13} that the diffusivity and the fast mobility are related by the law: 
\begin{equation}\label{DVsu2}
D =  M^{-\alpha} \mathcal{F}_\alpha(\langle u^2 \rangle )
\end{equation}
where $\mathcal{F}_\alpha$ is a state-independent function and $\alpha$ is equal to $0$ or $1$ in binary mixtures or fully-flexible, unentangled polymers, respectively. 
A qualitative understanding of Eq.\ref{DVsu2} is provided by the following argument. For atomic systems $D \sim \langle u^2 \rangle/ \tau_\alpha$, whereas for fully-flexible unentangled polymers $D \sim R^2_{ee}/ \tau_{ee} \sim (M-1) b^2/ (4 M^2  \tau_\alpha) \sim b^2/ (4 M \tau_\alpha)$ where $b$, $R_{ee}$ and $ \tau_{ee}$ are the bond length, the  end-end distance and the average reorientation time of the polymer chain, respectively \cite{DoiEdwards}. Reminding that $\langle u^2 \rangle$ is virtually independent of $M$ \cite{OurNatPhys} and resorting to Eq.\ref{eqn:u2tauExp}, we see that the previous approximated expressions of $D$ comply with  Eq.\ref{DVsu2}.
By resorting to Eq.\ref{eqn:u2tvg}, one concludes from Eq.\ref{DVsu2} that the $\gamma_{ts}$ exponent of the diffusivity must be equal to the one of the fast mobility (at least for unentangled polymers and binary atomic mixtures). Support to this conclusion is gained by considering the TS scaling of  decamers ($M=10$) with semi-rigid bond and interacting via the LJ  potential ($(p,q)=(6,12)$). The TS scaling of the {fast mobility}  occurs with $\gamma_{ts} = 6.7(1)$, see Table \ref{tab:u2tvg}, which is rather close to the TS exponent of the diffusivity, $\gamma_{ts} = 6$ \cite{budzienJCP04}. Virtual coincidence between the TS exponents of the fast mobility and the diffusivity is found in binary mixtures \cite{SpecialIssueJCP13}.}

\begin{figure}[t]
  \begin{center}
    \includegraphics[width=0.95\linewidth]{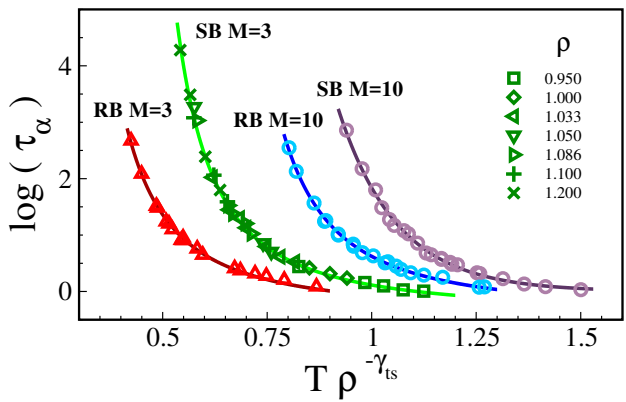}
    \caption[uno]{\label{fig:tauTvGamma} \small{TS of the structural relaxation time $\tau_\alpha$ from MD simulations of chains with length $M=3,10$, LJ non-bonding potential and rigid  (RB) or semi-rigid (SB) bonds. 
 The states span the ranges $0.95 \le \rho \le 1.2$, $0.5 \le T \le 1.0$. Details are found in Appendix A2 of 
ref. \cite{TesiFrancesco}. For illustrative purpose  the density of the states of the trimers with SB bonds is shown. 
For clarity reasons data are horizontally shifted: RB $M=3$ (shift=$+0.0$), SB $M=3$ (shift=$+0.2$), RB $M=10$ (shift=$+0.4$) and SB $M=10$ (shift=$+0.6$). The continuous line is Eq. \ref{eqn:tauTvgMD} with parameters from Table \ref{tab:u2tvg} and $\alpha =-0.424(1), \beta = 2.7(1) \cdot 10^{-2}, \gamma =  3.41(3) \cdot 10^{-3}$  \cite{OurNatPhys}. No adjustement is done. The TS exponent $\gamma_{ts}$ is equal to the corresponding one of the fast mobility. { All the quantities are in reduced MD units.}}}
  \end{center}
\end{figure}

\subsubsection{Thermodynamic scaling beyond the cage regime}
\label{mastercurveRes2}

The cage effect is {\it missing} if $\langle u^2 \rangle > \langle u^2_m \rangle= 0.125$, see Sec.\ref{MobilRelaxElastic}. We observe TS  of the fast mobility even if $\langle u^2 \rangle > 0.125$. This is shown by Figure \ref{fig:u2Tvglong} where one  observes that the scaling exponent $\gamma_{ts} = 5.80$ found in the cage regime collapses the fast-mobility on a TS master curve also for $\langle u^2 \rangle > \langle u^2_m \rangle$. The finding strongly suggests that the thermodynamic scaling does not rely on specific aspects of the supercooled regime. We also note that the linear form given by Eq. \ref{eqn:u2tvg} does not break down abruptly when the cage effect disappears at $\langle u^2_m \rangle$ but it provides a good approximation of the TS master curve up to, say, $\sim 2 \langle u^2_m \rangle$.

\begin{figure}[t]
  \begin{center}
    \includegraphics[width=0.99\linewidth]{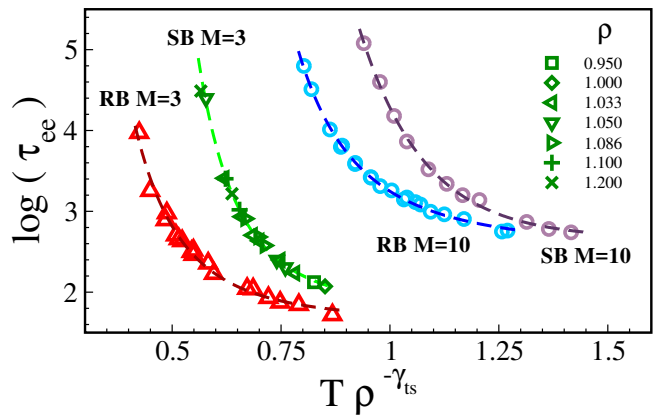}
    \caption[uno]{\label{fig:teeTvGamma} \small{TS of the chain reorientation time $\tau_{ee}$ for the same systems of Fig. \ref{fig:tauTvGamma}. For clarity reasons, data are horizontally shifted as in Fig. \ref{fig:tauTvGamma}. The dashed lines are guides for the eyes. { All the quantities are in reduced MD units.}}}
  \end{center}
\end{figure}

\subsection{Thermodynamic scaling of relaxation}
\label{TSRelax}

We now show that the MD simulations indicate that TS of the fast mobility with exponent $\gamma_{ts}$ also leads  to TS of the structural relaxation with the {\it same} exponent. To this aim, we recast Eq.\ref{eqn:u2tauExp} as
\begin{equation}
\log \tau_\alpha = {\alpha} + {\beta} \, \frac{1}{\langle u^2\rangle} + {\gamma} \, \frac{1}{\langle u^2\rangle^{2}}
\label{parabolaMD}
\end{equation}
For the present polymer model, irrespective of the non-bonding potential and the chain length, one has $\alpha =-0.424(1), \beta = 2.7(1) \cdot 10^{-2}, \gamma =  3.41(3) \cdot 10^{-3}$  \cite{OurNatPhys}. Plugging the TS linear master curve Eq.\ref{eqn:u2tvg} into Eq.\ref{parabolaMD}  gives:
\begin{equation}
\label{eqn:tauTvgMD}
\log \tau_\alpha=\alpha+\frac{\beta}{\left ( a_0 +  a_1\cdot T\rho^{-\gamma_{ts}} \right)}+\frac{\gamma}{\left ( a_0 +  a_1\cdot T\rho^{-\gamma_{ts}} \right)^2}
\end{equation}
with $a_0$, $a_1$ and $\gamma_{ts}$ as given by Table \ref{tab:u2tvg}.

Fig. \ref{fig:tauTvGamma} compares Eq. \ref{eqn:tauTvgMD} with the structural relaxation of a selected set of systems. We observe that: i) the TS exponent $\gamma_{ts}$ of the fast mobility also results in TS of the structural relaxation over about four decades of the relaxation time, ii) the TS master curve is well represented by Eq.\ref{eqn:tauTvgMD}.

To complete the analysis, we consider the average reorientation time $\tau_{ee}$ of the chain, i.e., the decay time of the correlations of the vector joining the end monomers of the chain \cite{lepoJCP09}. For fully-flexible unentangled linear polymers $\tau_{ee}$ increases as $M^2$, whereas $\tau_\alpha$ depends weakly on $M$ \cite{DoiEdwards}. Fig. \ref{fig:teeTvGamma} shows that the exponent $\gamma_{ts}$ of the fast mobility also provides TS of $\tau_{ee}$. The explicit form of the TS master curve of $\tau_{ee}$ is not given here. In fact, even there is a strong correlation between $\tau_{ee}$ and the fast mobility, the relation differs from Eq. \ref{parabolaMD} \cite{lepoJCP09}, so that we cannot extend Eq.\ref{eqn:tauTvgMD} to $\tau_{ee}$.
The coincidence of the scaling exponent for the segment relaxation  and the chain reorientation has been noted in poly(propylene glycol), 1,4-polyisoprene as well as in poly(oxybutylene) \cite{NgaiMM2005,RolandBook11}.  Nonetheless, Fragiadakis et al, investigating very carefully the density scaling in 1,4-Polyisoprene (PI) of different molecular weight by dielectric relaxation, noted that there is a small difference in the exponent $\gamma_{ts}$ for the segmental and the chain modes of the lowest molecular weight PI with degree of polymerization $18$ \cite{RolandFragiadakisMM2011}.

We conclude this section by stressing that, even if the coincidence of the TS exponent for different dynamical quantity was reported \cite{casaliniRepProg05}, it is of remarkable interest here that the {\it same} exponent $\gamma_{ts}$ is able to scale the picosecond fast mobility, the slow structural relaxation and the even slower chain reorientation up to diffusivity.

\begin{figure}[t]
  \begin{center}
    \includegraphics[width=0.8\linewidth]{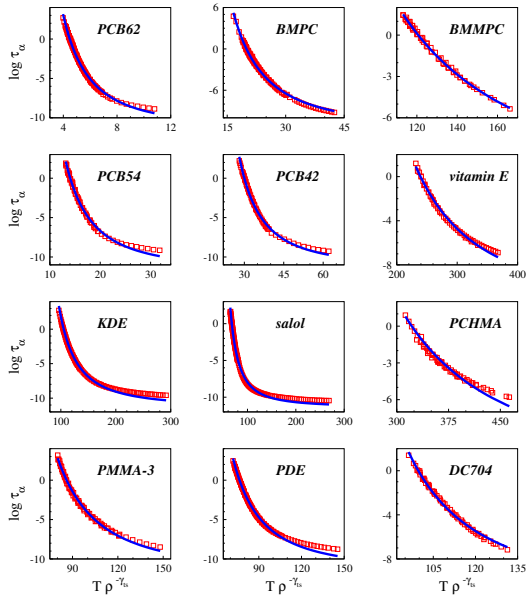}
    \caption[uno]{\label{buoni} TS, $\log \tau_\alpha$ vs $T \rho^{-\gamma_{ts}}$, of glassformers (squares) with lower isochoric fragility.  The best-fit with Eq.\ref{eqn:tvGexpL} is superimposed (continuous line). Best-fit values in Table \ref{tab:expData}. }
  \end{center}
\end{figure}

\begin{figure}[t]
  \begin{center}
    \includegraphics[width=0.8\linewidth]{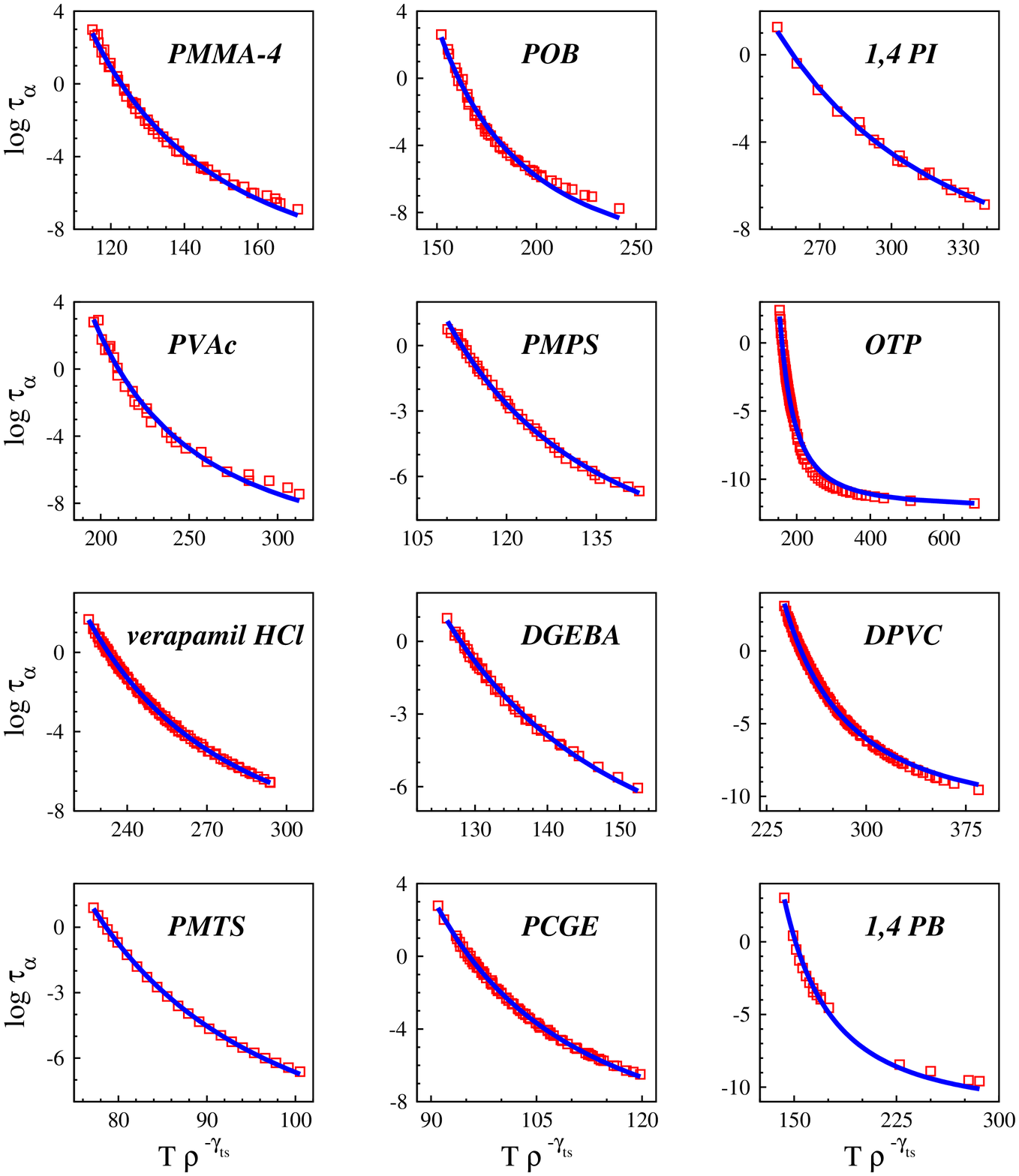}
    \caption[uno]{\label{medi} TS, $\log \tau_\alpha$ vs $T \rho^{-\gamma_{ts}}$, of glassformers (squares) with intermediate isochoric fragility.  The best-fit with Eq.\ref{eqn:tvGexpL} is superimposed (continuous line). Best-fit values in Table \ref{tab:expData}. }
  \end{center}
\end{figure}

\begin{figure}[t]
  \begin{center}
    \includegraphics[width=0.8\linewidth]{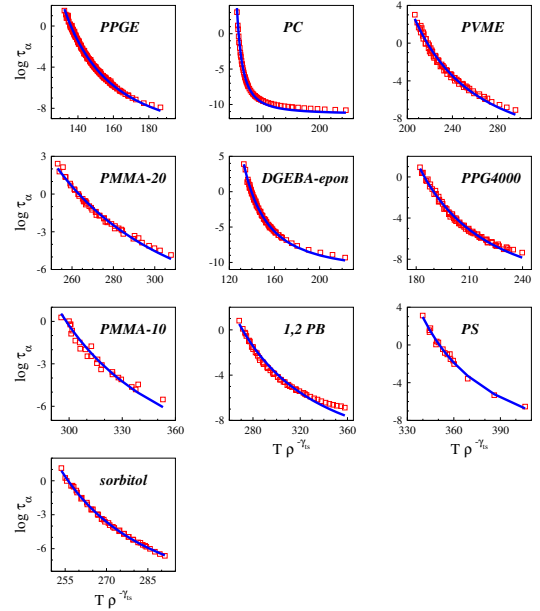}
    \caption[uno]{\label{cattivi} TS, $\log \tau_\alpha$ vs $T \rho^{-\gamma_{ts}}$, of glassformers (squares) with higher isochoric fragility.  The best-fit with Eq.\ref{eqn:tvGexpL} is superimposed (continuous line). Best-fit values in Table \ref{tab:expData}. }
  \end{center}
\end{figure}

\subsection{Comparison with the experiments}
The results of the MD investigation concerning TS of the fast mobility and relaxation pose the question whether they are limited to the specific class of scrutinised glassformers  or they capture general aspects of TS. To this aim, motivated by the findings of Sec.\ref{mastercurveRes} and Sec.\ref{TSRelax}, we now establish contact between the experimental and the MD results. First, we derive the TS master curve of the structural relaxation by combining the universal scaling Eq.\ref{eqn:u2tauExp} with the linear TS master curve of the fast mobility, Eq.\ref{eqn:u2tvg} recast as:
\begin{equation}
\label{eqn:u2tvg1}
\langle u^2\rangle=\langle u_g^2\rangle \left [ 1 + \kappa \left ( T\rho^{-\gamma_{ts}} - T_g\rho_g^{-\gamma_{ts}} \right ) \right ]
\end{equation}
where GT occurs when the scaling quantity $T\rho^{-\gamma_{ts}}$ is equal to  $T_g\rho_g^{-\gamma_{ts}}$. When applied to experimental data, it is understood that the quantity $\rho$ is the mass density and not the number density as in MD studies. Combining Eq. \ref{eqn:u2tauExp} and Eq. \ref{eqn:u2tvg1}, we obtain
\begin{equation}\label{eqn:tvGexpL}
\log \tau_\alpha=\alpha'+\frac{\tilde{\beta}}{\left [ 1 + \kappa \left ( Y - Y_g \right ) \right ]}+\frac{\tilde{\gamma}}{\left [ 1 + \kappa \left ( Y - Y_g \right ) \right ]^2}
\end{equation}
where
\begin{equation}
\label{defY}
 Y\equiv T\rho^{-\gamma_{ts}}
\end{equation}
\clearpage
\onecolumngrid
\begin{center}
\begin{table}[t]
 \begin{tabular}{| l | c | c | c | c | c | c | c | }
\hline	
  {} & {System} 	&  $\gamma_{ts}$ 	& $\kappa \cdot 10^2 $ 	& $\alpha'$ 	& $m_v^{fit}$  	& $m_v^{exp}$ & Ref. \\
  {} & {}		& {}			& {\footnotesize $[\mbox K^{-1}(g/cm^3)^{\gamma_{ts}}]$}	& {}		& {}			& {}			&	\\		
\hline
	1 	&	PCB62	& $8.5$	& 	$26.5$		& $-11.18$	& $28 \pm 2$	& $24 \pm 1$		&	\cite{RolandJCP05}			\\	
	2 	&	BMPC	& $7.0$  	&	$5.72$		& $-11.45$	& $28 \pm 2$	& $25 \pm 1$		&	\cite{paluchPRB02,HenselJCP02}			\\	
	3 	&	BMMPC	& $8.5$  	& 	$0.891$		& $-11.48$	& $26 \pm 2$	& $26 \pm 1$		&	\cite{paluchPRB02}			\\
	4	&  	PCB54	& $6.7$	& 	$10.23$		& $-11.49$	& $35 \pm 3 $	& $31 \pm 3$		&	\cite{RolandJCP05}			\\
	5	&  	PCB42	& $5.5$	& 	$6.16$		& $-11.12$	& $46 \pm 4$	& $35 \pm 5$		&	\cite{RolandJCP05}			\\
	6	&  	vitamin E	& $3.9$	& 	$0.52$		& $-11.92$	& $31 \pm 2 $	& $36 \pm 6$		&	\cite{kaminskiPREvitaE}		\\
	7	&  	KDE		& $4.5$	& 	$1.62$		& $-11.01$	& $42 \pm 3 $	& $39 \pm 3$		&	\cite{casaliniPRB05}			\\
	8	&  	salol		& $5.2$	& 	$3.16$		& $-11.00$	& $53 \pm 4 $	& $40 \pm 5$		&	\cite{salol_Roland}			\\
	9	&  	PCHMA	& $2.9$	& 	$0.373$		& $-12.00$	& $30 \pm 2 $	& $42 \pm 8$		&	\cite{rolandMM07}			\\
	10	&  	PMMA-3	& $3.7$	& 	$1.93$		& $-11.65$	& $42 \pm 3 $	& $43 \pm 2$		&	\cite{capacciPMMA}			\\
	11	&  	PDE		& $4.4$	&	$2.58$		& $-11.26$	& $49 \pm 3 $	& $45 \pm 4$		&	\cite{paluchPRB02,PaluchJCP02,casaliniPRB05} 			\\
	12 	&	DC704	& $6.15$	&	$2.26$		& $-11.10$	& $56 \pm 4 $	& $47 \pm 5$		&	\cite{GundermannNatPhys11}	\\	
	13	&  	PMMA-4	& $3.2$	&	$1.42$		& $-11.70$	& $44 \pm 3 $	& $49 \pm 3$		&	\cite{capacciPMMA}			\\
	14	&  	POB		& $2.65$	&	$1.15$		& $-11.77$	& $47 \pm 4 $	& $50 \pm 6$		&	\cite{casaliniMMpob05}		\\
	15 	&	1,4 PI	& $3.5$	& 	$0.714$		& $-11.92$	& $47 \pm 4 $	& $51 \pm 7$		&	\cite{FloudasJCP99,RolandJPSB04,casaliniRepProg05}			\\	
	16	&  	PVAc	& $2.6$	&	$0.870$		& $-11.39$	& $46 \pm 5 $	& $52 \pm 5$		&	\cite{rolandMMpvac03}		\\
	17 	&	PMPS	& $5.63$	& 	$2.16$		& $-11.37$	& $61 \pm 4 $	& $54 \pm 3$		&	\cite{paluchJCPpmps02}		\\	
	18 	&	OTP		& $4.0$	& 	$1.30$		& $-11.74$	& $53 \pm 4 $	& $54 \pm 2$		&	\cite{NaokiJPC87,NaokiJPC89,HansenStickelJCP97,ZamponiJPCB08,tolleRepProg01,TollePRL98}			\\	
	19 	&	verapamil HCl	& $2.47$	& 	$0.969$		& $-11.50$	& $57 \pm 3 $	& $57 \pm 3$		&	\cite{Dlubek_EJPS10,WojnarowskaJCP10}		\\	
	20	&	DGEBA	& $2.8$	&	$2.08$		& $-12.49$	& $68 \pm 5 $	& $57 \pm 7$		&	\cite{PaluchDGEBA03}		\\
	
	21 	&	DPVC	& $3.2$	& 	$1.04$		& $-11.42$	& $66 \pm 4 $	& $62 \pm 3$		&	\cite{CapaccioliJNCS09}		\\	
	22 	&	PMTS	& $5.0$	& 	$2.69$		& $-11.71$	& $54 \pm 3 $	& $63 \pm 2$		&	\cite{paluchMMpmts}			\\	
	23	&	PCGE	& $3.3$	& 	$2.52$		& $-11.38$	& $61 \pm 4 $	& $63 \pm 3$		&	\cite{rolandComJCP04} 		\\
	24	&  	1,4 PB	& $1.8$	& 	$1.45$		& $-11.56$	& $55 \pm 7 $	& $64 \pm 6$		&	\cite{simionescoEPL04}		\\
	25	&  	PPGE	& $3.45$	&	$2.01$		& $-11.36$	& $69 \pm 4 $	& $65 \pm 4$		&	\cite{CasaliniPRE01,CorezziJCP02}			\\
	26	&	PC		& $3.8$	& 	$5.04$		& $-11.00$	& $72 \pm 5 $	& $66 \pm 4$		&	\cite{pawlusPRE04}			\\	
	27	&  	PVME	& $2.5$	& 	$0.923$		& $-11.91$	& $51 \pm 4 $	& $66 \pm 7$		&	\cite{casaliniJCPpvme}		\\
	28	&	PMMA-20 & $1.94$	&	$0.818$		& $-11.86$	& $55 \pm 4 $	& $67 \pm 13$		&	\cite{CasaliniPMMA13}		\\	
	29	&  	DGEBA-epon	& $3.5$	& 	$2.19$		& $-11.49$	& $78 \pm 5 $	& $70 \pm 8$	&	\cite{NgaiJCP12}			\\
	30	&  	PPG4000	& $2.5$	& 	$1.41$		& $-11.93$	& $67 \pm 5 $	& $76 \pm 15$		&	\cite{rolandPPG4000}		\\
	31	&  	PMMA-10	& $1.8$	&	$0.845$		& $-12.00$	& $65 \pm 8 $	& $85 \pm 20$		&	\cite{capacciPMMA}			\\
	32	&  	1,2 PB	& $1.89$	& 	$0.817$		& $-12.00$	& $56 \pm 4 $	& $86 \pm 15$		&	\cite{casaliniPRB05} 			\\	
	33 	&	PS		& $2.27$	&	$1.125$		& $-11.46$	& $101 \pm 10 $	& $104 \pm 8$		&	\cite{GuoJCP11}			\\	
	34 	&	sorbitol	& $0.18$	&	$1.65$		& $-11.62$	& $108  \pm 8 $	& $112 \pm 10$		&	\cite{HenselPRL02,casaliniPRE04,CasaliniGamacheJCP11}			\\	
	\hline
 \end{tabular}
\caption{\label{tab:expData} Best-fit values of the parameters of the TS master curve Eq.\ref{eqn:tvGexpL} ($\kappa$ and $\alpha'$, adjusted in the range $\alpha - 0.5 \le \alpha' \le \alpha + 0.5$ with $\tilde{\beta} = 1.62(6)$ and $\tilde{\gamma} = 12.3(1)$ \cite{OurNatPhys} ) for the glassformers in Fig.\ref{buoni}, Fig.\ref{medi}, Fig.\ref{cattivi}.
The experimental characteristic exponent $\gamma_{ts}$, the isochoric fragility $m_v^{exp}$, Eq.\ref{eqn:fragVdeff}, and the best-fit value $m_v^{fit}$, evaluated via  Eq.\ref{eqn:fragMC}, are also listed. The glassformers are listed in increasing order of the experimental  isochoric fragility $m_v^{exp}$.
}
\end{table}
\end{center}
\twocolumngrid
\noindent
 In Eq.\ref{eqn:tvGexpL}  rigorously $ \alpha' = \alpha = 2 - \tilde{\beta} - \tilde{\gamma}$ with the definition $\log \tau_\alpha = 2$ where $\tilde{\beta} = 1.62(6)$ and $\tilde{\gamma} = 12.3(1)$ are universal values, independent of the system \cite{OurNatPhys}. However, we consider $\alpha'$ as mildly adjustable in the range  $\alpha - 0.5 \le \alpha' \le \alpha + 0.5$ to account for small errors in the determination of the glass transition. Taking $Y_g$ from the experiment, the total number 
of adjustable parameters of Eq.\ref{eqn:tvGexpL} is two ($\alpha'$ and the slope $\kappa$),  which is less than the number, three, of alternative expression of the TS master curve of the structural relaxation  \cite{CasaliniAvra06,CasaliniJNCS07,PaluchSokolovTvgammaMauroJPCLett12}.
 
 Figs. \ref{buoni}, \ref{medi} and \ref{cattivi}  show the comparison of Eq. \ref{eqn:tvGexpL} with the TS master curves of the structural relaxation of thirty-four different glassformers spanning a large range of the scaling exponent which controls the density influence on relaxation ($0.18 \lesssim \gamma_{ts} \lesssim 8.5$). The best-fit parameters are listed in Table \ref{tab:expData}.   Despite Eq. \ref{eqn:tvGexpL} has only two adjustable parameters - with narrowly bounded $\alpha'$ -, it  provides, all in all, an effective analytical expression of the TS master curve over a wide range of relaxation times, e.g. about fourteen decades for BMPC, see Fig.\ref{buoni}, and the prototypical glassformer OTP, see Fig.\ref{medi}. Nonetheless, deviations are seen, especially for short relaxation times. {In principle, the deviations could be ascribed to the limited accuracy of Eq. \ref{eqn:tvGexpL}  for states with weak cage effect. However, for some glassformers, e.g. PCHMA, PDE, POB, 1,2 PB, deviations are apparent already for $T \rho^{-\gamma_{ts}}/ T_g  \rho_g^{-\gamma_{ts}} \gtrsim 1.35$ and  
 $\tau_\alpha \lesssim 10^{-6}-10^{-8}$ where the above argument is untenable, so, to date we are unable to reach a clear conclusion about the issue. 
Since Eq. \ref{eqn:tvGexpL} relies on Eq. \ref{eqn:u2tauExp}, one could think that the latter breaks down at short relaxation times. However, we know that  experimental data
 validated Eq. \ref{eqn:u2tauExp}  down to about $10-100$ picoseconds \cite{OurNatPhys,UnivPhilMag11}. Then, we believe that deviations follow by the other relation leading to Eq.\ref{eqn:tvGexpL}, i.e. the TS master curve of the fast mobility, Eq.\ref{eqn:u2tvg1}, which apparently must be improved at high $T \rho^{-\gamma_{ts}}$ values.

 It is noteworthy that the deviations occur for times much shorter than those where the so-called dynamic crossover has been observed for the same systems \cite{casaliniPRB05}. In oher words, Eq. \ref{eqn:tvGexpL} is able to represent data across the time range where a break of the VFT has been observed, but it fails at much shorter times.  Other fitting funtions, like that based on Avramov model, have an additional parameter, often taking into account anharmonicity of the potential. Thanks to this additional parameters they can provide a better scaling. Further experimental and numerical studies will be carried out with the aim to test the range of validity of Eq. \ref{eqn:tvGexpL}. 

The best-fit of Eq. \ref{eqn:tvGexpL} does not deviate from the experimental TS master curves close to GT. This is worth noting since Eq. \ref{eqn:tvGexpL} relies on Eq.\ref{eqn:u2tvg1} which fails at small $T\rho^{-\gamma_{ts}}$ values, see Sec.\ref{mastercurveRes}. By reminding that, for a given density $\rho_0$, the temperature $T_{0}^{(FM)}$ where $\langle u^2(T)\rangle $ vanishes has been associated to the the Vogel-Fulcher-Tammann (VFT) temperature  $T_0$ - which is lower than $T_g$ - \cite{douglasPNAS09}, we speculate that around GT the linear approximation of the TS master curve of the fast mobility Eq.\ref{eqn:u2tvg1} is still reliable.

\begin{figure}[t]
  \begin{center}
    \includegraphics[width=0.9\linewidth]{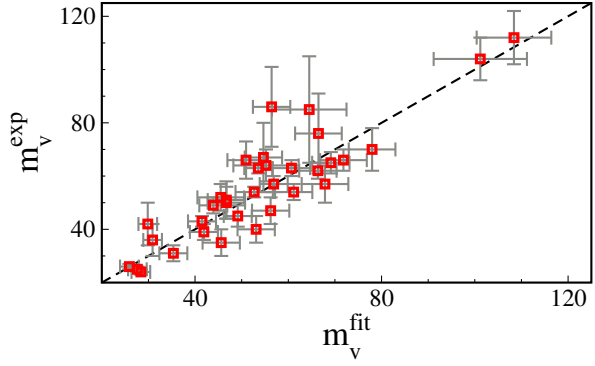}
    \caption[uno]{\label{fig:frag} Correlation plot between the best-fit value of the isochoric fragilities evaluated according to Eq.\ref{eqn:fragMC} and the experimental one (Pearson correlation coefficient $R =0.89$). The dashed line is the bisectrix.}
  \end{center}
\end{figure}

\begin{figure}[t]
  \begin{center}
    \includegraphics[width=0.99\linewidth]{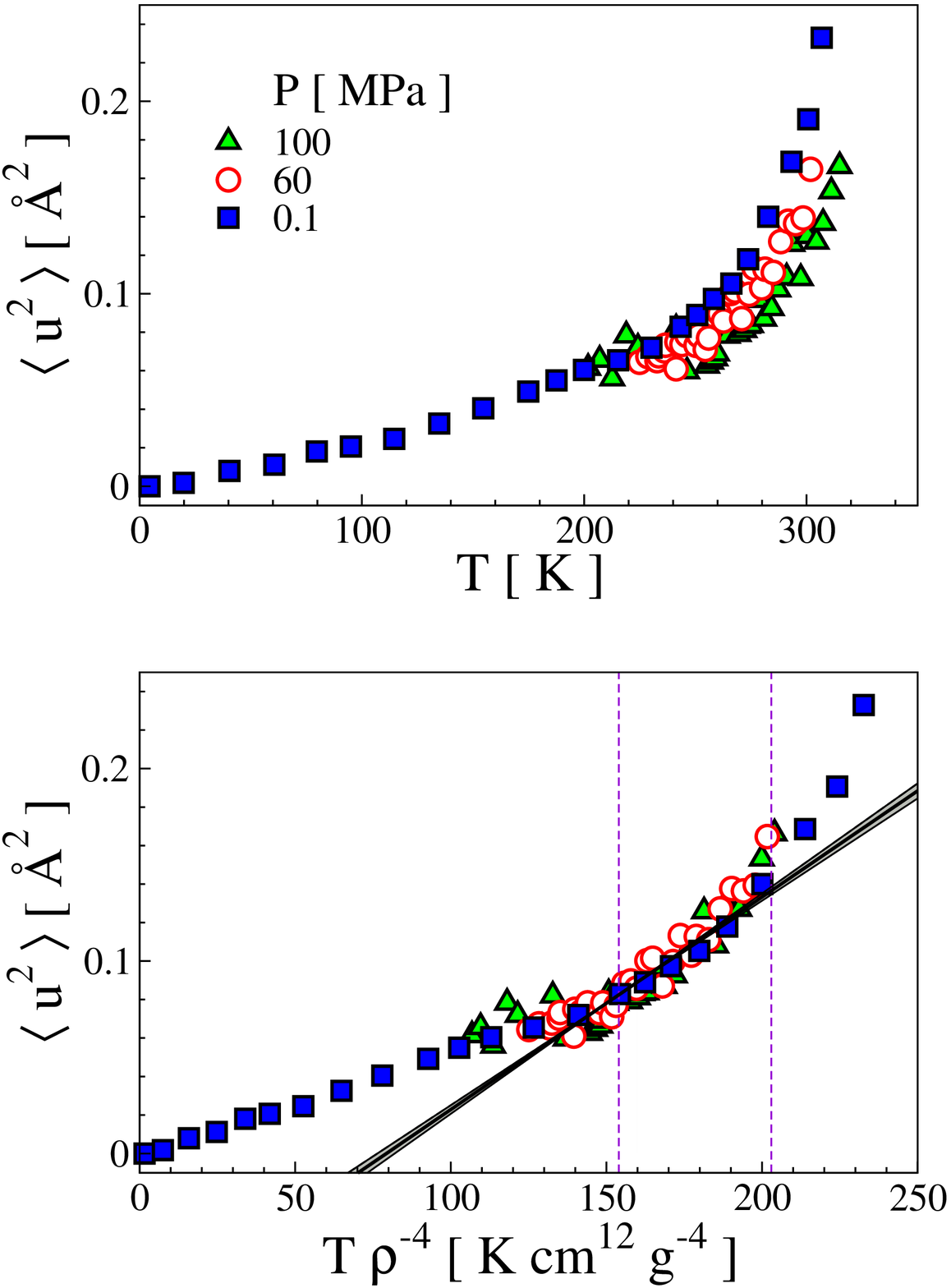}
    \caption[uno]{\label{fig:u2OTP} Top: fast mobility of OTP from incoherent elastic scattering intensity $S(Q,0)$ as a function of temperature for different pressures \cite{tolleRepProg01}. Bottom: TS of the data with $\gamma_{ts} =4$ (equation of state from ref.\cite{NaokiJPC89}). 
{    The vertical dashed lines mark the glass transition ($\tau_\alpha = 10^2$ s, $T_g\rho_g^{-\gamma_{ts}} = 154$ K cm${}^{12}$ g${}^{-4}$, $\langle u^2_g \rangle = 0.0829$ \AA ${}^2$),  and the critical value $T_c\rho_c^{-\gamma_{ts}}$ above which the quasielastic broadening from structural relaxation  contribution, being $\tau_\alpha=20$ ns, has to be taken into account \cite{tolleRepProg01} and the measured $\langle u^2 \rangle$ cannot be longer interpreted as the fast mobility within the cage.
   The thick solid line is Eq.\ref{eqn:u2tvg1} with $\kappa$ evaluated from Eq.\ref{eqn:fragMC} with $\tilde{m}_v = m_v^{exp}$, as taken from Table \ref{tab:expData}. The two thin black lines bound the uncertainty on the slope due to the one on $m_v^{exp}$.}
}
  \end{center}
\end{figure}

To provide further evidence about the accuracy of Eq. \ref{eqn:tvGexpL} close to GT, we now consider the isochoric fragility $m_v$, { namely the slope of the master curve of the structural relaxation, Eq.\ref{eqn:tvgammdef}, at the glass transition, which in terms of Eq.\ref{defY} is given by}:
\begin{equation}
\label{eqn:fragVdeff}
m_v=\left .  \frac{\partial\log\tau_\alpha}{\partial\left (Y_g/Y\right)}\right |_{Y_g}
\end{equation}
Plugging Eq. \ref{eqn:tvGexpL} into Eq.\ref{eqn:fragVdeff} leads to:
\begin{equation}\label{eqn:fragMC}
\tilde{m}_v=\kappa \left ( \tilde{\beta}+2\tilde{\gamma} \right) T_g\rho_g^{-\gamma} 
\end{equation}
The comparisons between $m_v^{fit}$, as taken from Eq.\ref{eqn:fragMC} with $\kappa$ from Table \ref{tab:expData}, and the experimental isochoric fragility, $m_v^{exp}$, as taken from Eq.\ref{eqn:fragVdeff}, is listed in Table \ref{tab:expData} and plotted in Fig.\ref{fig:frag}. It is seen that, apart from a few outliers, the correlation is rather good {, given the experimental uncertainties}. This, again, suggests that Eq.\ref{eqn:tvGexpL} is an effective TS master curve of the structural relaxation also close to GT.

An attempt to test Eq.\ref{eqn:u2tvg1}, stating the approximately linear character of the TS master curve of the fast mobility above GT, is presented in Fig. \ref{fig:u2OTP} for the prototypical glassformer OTP. { Fig. \ref{fig:u2OTP} (top) shows the pressure dependence of the fast mobility of OTP \cite{tolleRepProg01}. The TS scaling in the supercooled regime 
with the  {\it same} characteristic exponent $\gamma_{ts}$ of the TS master curve of the structural relaxation, see Table \ref{tab:expData}, is in Fig. \ref{fig:u2OTP} (bottom). The resulting master curve is 
compared to Eq.\ref{eqn:u2tvg1} with {\it no} adjustable parameters. In fact, we take $T_g\rho_g^{-\gamma_{ts}}$ from the OTP master curve in Fig. \ref{medi} where $\tau_\alpha = 10^2$ s and evaluate 
 the corresponding $\langle u^2_g \rangle$ from Fig. \ref{fig:u2OTP} (top). The slope $\kappa$ is evaluated from Eq.\ref{eqn:fragMC} by setting  $\tilde{m}_v = m_v^{exp}$, where $m_v^{exp}$ is the experimental value of the isochoric fragility, see Table \ref{tab:expData}.} 
 The agreement, even in the presence of some concavity of the experimental master curve, is quite satisfactory across the supercooled regime  down to the GT. It suggests that the TS master curve of the fast mobility is effectively approximated by a linear law in $T \rho^{-\gamma_{ts}}$ in the supercooled regime. { Notice that, as a matter of fact, the above procedure  predicts the TS master curve of the fast mobility on the sole basis of the experimental value of the isochoric fragility $m_v^{exp}$, see Table \ref{tab:expData}.}

\section{Conclusions}
\label{conclu}

{ The present paper investigates the thermodynamic scaling of the fast vibrational dynamics. In particular, we address ourselves to the fast mobility, the mean square amplitude of the picosecond rattling motion inside the cage, which is studied by extensive MD simulations and comparison with experimental results.  The MD simulations are carried out on a variety of coarse-grained polymer models of a melt of unentangled linear chains where both the bonding and non-bonding potentials as well as the chain length are changed. The polymer model with semi-rigid bonds exhibits TS with weak virial-energy correlations. This precludes the usual TS interpretation in terms of an effective inverse-power law potential replacing the actual particle-particle interaction potential.}
One major result of the MD simulations is the evidence of the joint TS with the {\it same}  characteristic exponent $\gamma_{ts}$ of both the fast mobility $\langle u^2 \rangle$ and the much slower structural relaxation and chain reorientation. 
We find that the TS master curve of the fast mobility in the cage regime is well described by a simple linear relation in $T\rho^{-\gamma_{ts}}$ with slope $\kappa$. The linear TS master curve is expected to be sufficiently accurate at GT and extends also in part of the liquid region where no caging is apparent, suggesting that TS is not related to supercooling. The linear master curves intersect nearly in a single point so that they can be approximately labelled by { their slope which is strictly related to the isochoric fragility}. By combining the linear TS master curve of the fast mobility with the universal relation linking the latter to the structural relaxation, we derive an analytical expression of the TS master curve of the structural relaxation with two adjustable parameters, one being narrowly  bounded and the other one being the slope $\kappa$. The theoretical TS master curve of the structural relaxation is compared with the experimental ones of thirty-four glassformers. It shows good accuracy, especially close to GT, as confirmed by the good correlations between the best-fit and the experimental isochoric fragility in the range $24 \le m_v \le 112$. { For the glassformer OTP  the isochoric fragility allows to satisfactorily predict  the TS master curve of the fast mobility with no adjustments.}

\begin{acknowledgments}
A generous grant of computing time from IT Center, University of Pisa and Dell${}^\circledR$ Italia is gratefully acknowledged.
\end{acknowledgments}

\bibliography{biblioTVgammaU2.bib}

\end{document}